\documentclass[journal]{IEEEtran}
\usepackage{amsfonts}
\IEEEoverridecommandlockouts

\ifCLASSINFOpdf
\else
\fi
\usepackage{epstopdf}
\usepackage{caption}
\usepackage{multirow}
\usepackage{cite}
\usepackage{stfloats}
\usepackage{color}
\usepackage{epsfig}
\usepackage{graphicx}
\usepackage{psfig}
\usepackage{epsf}
\usepackage{slashbox}
\usepackage{float}
\usepackage{amssymb }
\usepackage{cite}
\usepackage{stfloats}
\usepackage{color}
\usepackage{epsfig}
\usepackage{graphicx}
\usepackage{psfig}
\usepackage{epsf}
\usepackage{slashbox}
\usepackage{float}
\usepackage[cmex10]{amsmath}
\usepackage{algorithm}
\usepackage{algorithmic}
\begin{document}
\title{On the Performance and Optimization  for MEC Networks Using Uplink NOMA}
\author{%\IEEEauthorblockN{Y. Ye, G. Lu, R. Q. Hu, L.Shi}
\IEEEauthorblockN{Yinghui Ye$^1$, Guangyue Lu$^2$, Rose Qingyang Hu$^3$, Liqin Shi$^1$}\\
\IEEEauthorblockA{$^1$School of Telecommunications Engineering, Xidian University, China.\\
$^2$ Shaanxi Key Laboratory of Information Communication Network and Security,\\ Xi'an University of Posts and Telecommunications. %Shaanxi information communication network and security Laboratory, \\ Department of Electrical and Computer Engineering,
$^3$ Department of ECE, Utah State University, U.S.A.}\vspace*{-1em}
\thanks{  Personal use of this material is permitted.
However, permission to use this material for any other purposes must be
obtained from the IEEE by sending a request to pubs-permissions@ieee.org.}
\thanks{This work was supported by  the scholarship from China Scholarship
Council, the Natural Science Foundation of China  (61801382), the Science and Technology Innovation Team of Shaanxi Province for Broadband Wireless and Application (2017KCT-30-02), the US National Science Foundations grants under the grants NeTS-1423348 and the EARS-1547312. The corresponding author is Liqin Shi (liqinshi@hotmail.com).
}
}
%\author{\IEEEauthorblockN{Yinghui Ye, Guangyue Lu, Rose Qingyang Hu,~\IEEEmembership{Senior Member,~IEEE}, Liqin Shi}
%\thanks{
%Yinghui Ye and Liqin Shi  are with the State Key Laboratory of Integrated Service Networks Lab, Xidian University, Xi'an, China (e-mail: connectyyh@126.com, liqinshi@hotmail.com).
%}
%\thanks{
%Guangyue Lu is with the Shaanxi Key Laboratory of Information Communication Network and Security, Xi'an University of Posts \& Telecommunications, China (e-mail: tonylugy@163.com).
%}
%\thanks{ Rose Qingyang Hu is with the Department of Electrical and Computer Engineering at Utah State University, U.S.A (e-mail:  rosehu@ieee.org).
%}
%\thanks{The research reported in this article was supported by the Natural Science Foundation of China (61801382), the scholarship from China Scholarship
%Council, the Science and Technology Innovation Team of Shaanxi Province for Broadband Wireless and Application (2017KCT-30-02), the 111 Project of China (B08038), the US National Science Foundations grants under the grants NeTS-1423348 and the EARS-1547312.}
%}This paper has been accepted by IEEE ICC Workshop 2019.
\markboth{ }
{Ye\MakeLowercase{\textit{et al.}}: Time Switching Enabled DF relaying with Direct Link:  Improved Relaying Protocols}
\maketitle

\begin{abstract}
In this paper, we investigate a non-orthogonal multiple access (NOMA) based mobile edge computing (MEC) network, in which two users may partially offload their respective tasks to a single MEC server through uplink NOMA. We propose a new offloading scheme that can operate in three different modes, namely the partial computation offloading, the complete
 local computation, and the complete offloading. We further derive a closed-form expression of the successful computation probability for the proposed scheme. As part of the  proposed offloading scheme, we formulate a problem to maximize the successful computation probability by jointly optimizing the time for offloading, the power allocation of the two users and the offloading ratios which decide how many tasks should be offloaded to the MEC server. We obtain the optimal solutions in  the closed forms. Simulation results show that our proposed scheme can achieve the highest successful computation probability than the existing schemes.
\end{abstract}
%\begin{IEEEkeywords}
%NOMA, mobile edge computing, offloading scheme, successful computation probability.
%\end{IEEEkeywords}
\IEEEpeerreviewmaketitle
\section{Introduction}
Mobile edge computing (MEC) has been deemed a promising technique to enhance computation service so that
future wireless communications are able to realize  computation-intensive and delay-sensitive services, e.g., virtual reality and autonomous driving \cite{8016573, 8434285,8611399}.  The basic idea of MEC is to  let mobile users perform
computation offloading, i.e., mobile users can offload  partial or complete  tasks to the nearby access points with more powerful computation capabilities.   There are two operation modes for MEC:   binary computation offloading and partial computation offloading \cite{8016573}. For the former operation mode, the computation tasks are  either fully locally computed or completely  offloaded to the MEC server. For the latter one, the computation tasks can be divided into two parts, where one part is locally executed and the other part is offloaded to the MEC server.

On the other hand, non-orthogonal multiple access (NOMA) has been viewed as a key technology for the future wireless communication networks \cite{8357810,7972929}. It allows  multiple users to operate in the same frequency band simultaneously with different power levels to improve the spectral efficiency and connectivity. By employing NOMA technology instead of  orthogonal multiple
access (OMA) to offload tasks, the offloading latency can be reduced and the performance of MEC can be improved. Therefore, the aforementioned two communication techniques, MEC and NOMA, can be combined  to obtain gains in terms of latency \cite{8492422}, energy consumption \cite{8537962, 8267072}. %Until now, several works have revealed the benefits  for the application of NOMA to MEC from the latency's and energy consumptions' perspective.% and attracted  extensive attentions recently. , yielding a new concept dubbed NOMA based MEC

The combination of NOMA and MEC was  studied in \cite{8537962}, where the authors minimized the weighted sum of the energy consumption at all mobile users
subject to their computation latency constraints for both the partial computation offloading  and the  binary computation offloading modes. A similar problem was  studied in \cite{8267072} by  considering the user clustering for uplink NOMA. In \cite{849242A}, the authors provided a guideline to choose the best mode among OMA, pure NOMA, and hybrid NOMA based MEC networks in terms of the energy consumed by complete offloading. Different from the previous works, where the main focus was on the energy computation minimization by optimizing the network parameters based on instantaneous channel state information (CSI), the authors in \cite{8467377} studied the impact of NOMA's parameters, e.g., user channel conditions and transmit powers, on the complete offloading by deriving the expression of the successful computation probability. To the best of our knowledge, there is no open work to study the successful computation probability of NOMA based  MEC networks for a hybrid operation mode, which takes   the  partial computation offloading, the full local computation and the complete offloading into consideration, based on the statistic CSI.

 \begin{figure}
    \centering
  \includegraphics[width=0.38\textwidth]{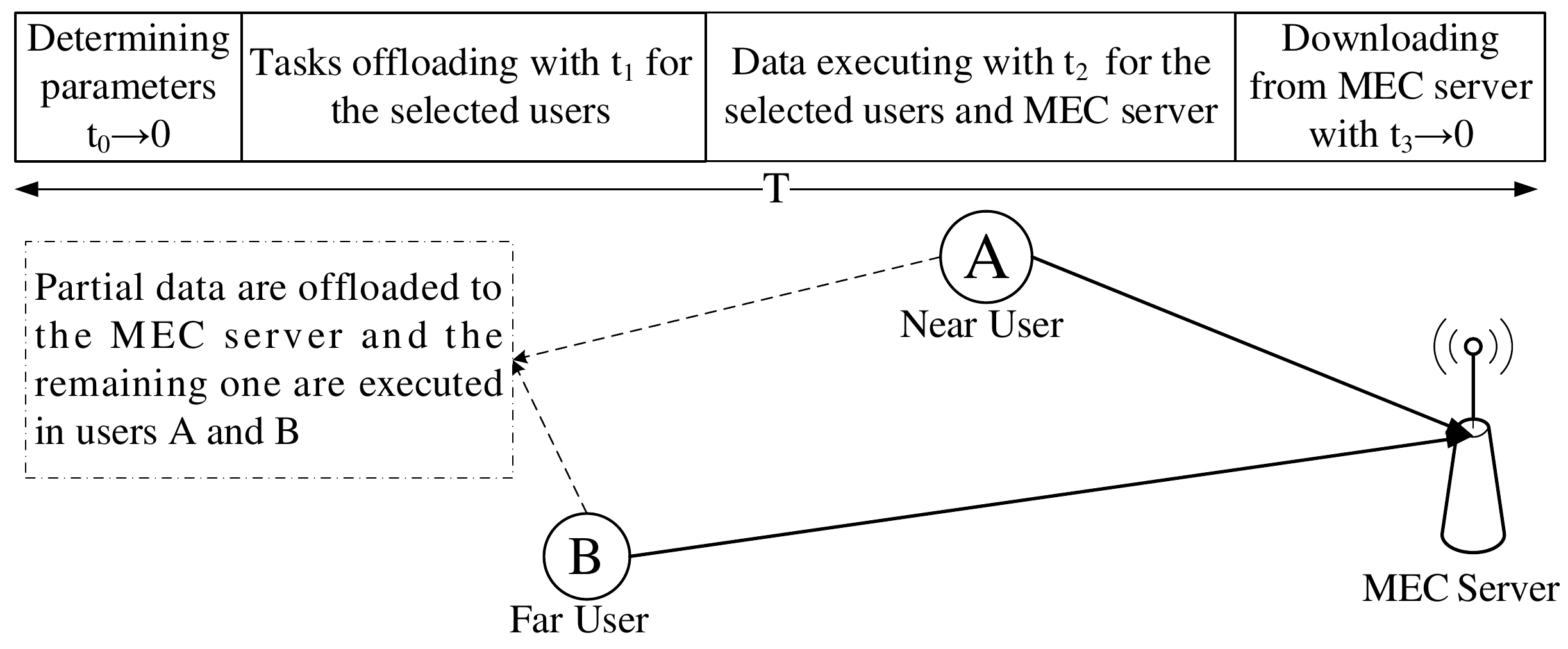}\\
  \caption{System Model.}
 \vspace{-15pt}
\end{figure}
In this paper, we consider a  NOMA based MEC network, where two users  may offload  their computation tasks to a single MEC server through uplink NOMA.  The proposed offloading scheme can  operate in one of following three modes, namely partial computation offloading,  complete local computation, and complete offloading. As part of the proposed offloading scheme,   we firstly derive a closed-form expression for the successful computation probability based on the statistic CSI. Then a problem is formulated to maximize the successful computation by optimizing the time for offloading, the power allocation of the two users to perform uplink NOMA and the offloading ratios which decide how many tasks should be offloaded to the MEC server. Although this problem is non-convex, we  obtain the optimal solutions in closed forms. Simulations are provided to support our work.
%Different from the existing works operating in a binary computation offloading or partial computation offloading,  in this paper, we propose a hybrid operation mode, which takes  both offloading modes into consideration.
%To the best of our knowledge, there is no open work to study the successful computation probability of NOMA based  MEC networks for a hybrid operation mode, where we takes   the  partial computation offloading, the fully local computation and the complete offloading into consideration, based on statistic CSI.
\section{System Model and Working Folw}
We consider a NOMA based MEC network consisting of one MEC server (i.e., gateway or base state) and two users $\rm{A}$ and $\rm{B}$, as shown in Fig. 1.  Each user has  tasks with $M_k$ ($k={\rm{A}}$, ${\rm{B}}$) bits to be executed  and the users may not be able to execute  their tasks locally within the latency budget due to the limited local computational capabilities. We assume that the task-input bits are bit-wise independent and can be arbitrarily divided into different groups \cite{8016573}. More specifically, for the user $k$, $\beta_k M_{{k}}$ bits are offloaded to the MEC server and the remaining ones, ${\left( {1 - {\beta_k }} \right){M_{{k}}}}$,  are locally executed, where $0\le \beta_k\le1$.
%For example, $\beta_k=1$ and $\beta_k=0$  correspond to the complete offloading and the fully local computation  at the  user $k$, respectively.
Therefore, the MEC server may schedule users to offload partial tasks  through uplink NOMA  so that all the tasks can be computed within the delay budget. We propose a new offloading scheme\footnote{The existing works \cite{8434285, 8537962, 8267072} assume that the users are able to offload   and execute tasks  simultaneously. Actually, the users are usually with limited computational capabilities, inferring that they may  be with   a  single core CPU. In this case, the assumption above may not hold  since a  single core CPU  can not execute more than one task (thread) simultaneously \cite{akhter2006multi}. Accordingly, in our proposed scheme, we assume that the users are installed with a single core CPU and switch the operation modes between  tasks offloading and tasks  executing. This is one of  main differences compared with the existing schemes.} that  operates in a hybrid operation mode, which can support  all three modes, namely  partial computation offloading, full local computation and  complete offloading. Moreover, all the channels are assumed to be quasi-static Rayleigh fading, where the channel coefficients are constant for each  block  but vary independently between different blocks. Also, the maximum supportable latency of the NOMA based MEC network is assumed as $T$ seconds and  $T$ is considered to be less than the coherence interval.

The proposed offloading scheme is introduced as follows. In  the first phase $t_0$, the MEC server determines the  parameters of the proposed offloading scheme such as  the offloading ratios and the time for offloading.  Then  users $\rm{A}$ and $\rm{B}$ offload their tasks to the MEC server via uplink NOMA at the second phase $t_1$. After successful offloading, the offloaded   and the local tasks are computed  at the MEC server and the users respectively during  the third phase $t_2$. Finally, the MEC server feeds back the computed   results  to  $\rm{A}$ and $\rm{B}$ within $t_3$. Following \cite{8537962, 8267072,8492422,8467377}, $t_0$ and  $t_3$,  are assumed very small and thus are neglected. Accordingly, our proposed offloading scheme consists of two main phases: offloading phase $t_1$ and   tasks  executing   phase $t_2$, i.e., $t_1+t_2\le T$.
\subsubsection{ Offloading Phase}
During the  offloading phase, $\rm{A}$ and $\rm{B}$ transmit their respective tasks with $\beta_{\rm{A}}M_{{\rm{A}}}$ and $\beta_{\rm{B}}M_{{\rm{B}}}$ bits   to the MEC server simultaneously.
The received signal at the MEC server is given as
\begin{small}
\begin{align}
y_{{\rm{MEC}}}^{{\rm{up}}} = \sqrt {\frac{{{P_{{{\rm{A}}}}}}}{{1 + d_{{{\rm{A}}}}^\alpha }}} {h_{{{\rm{A}}}}}{x_{{{\rm{A}}}}} + \sqrt {\frac{{{P_{{{\rm{B}}}}}}}{{1 + d_{{{\rm{B}}}}^\alpha }}} {h_{{{\rm{B}}}}}{x_{{{\rm{B}}}}} + w,
\end{align}
\end{small}where $P_{k}$ is the transmit power of user $k$; ${\frac{1}{{1 + d_{{k}}^\alpha }}}$ denotes the large-scale fading with the distance $d_{{{k}}}$ from the MEC server to the user $k$ and the path loss exponent $\alpha$; ${h_{{k}}} \sim {\mathbb{CN}}\left( {0,1} \right)$ models the small-scale Rayleigh fading between the MEC server and the user $k$;
${x_{{{k}}}}$ is the transmit signal at the user $k$ with $\mathbb{E}\left[ {{{\left| {{x_{{k}}}} \right|}^2}} \right] = 1$; $w$ is the received additive Gaussian white noise  at the MEC server with variance ${\sigma ^2}$. %${{P_{{{\rm{A}}}}}}=\lambda P$ and ${{P_{{{\rm{B}}}}}}=(1-\lambda) P$.

{\normalsize{In order to improve the performance of the uplink NOMA, we introduce the power allocation coefficient at the users, denoted by $\lambda$, to realize the power control at the users. Let ${{P_{{{\rm{A}}}}}}=\lambda P$ and ${{P_{{{\rm{B}}}}}}=(1-\lambda) P$ \cite{7542118}, where $P$ is the total power of the  two users\footnote{In many practical scenarios, one crucial criteria is the total transmission power. When multiple users share the
same bandwidth in a cell,  the constraint of the total transmission power  is important to manage inter-cell interference. Besides, as pointed out in \cite{7433470}, the total power constraint is beneficial to measure  the inter-group interference.  These facts  motivate us to consider this model, where the total  power of NOMA users is no more than a threshold,  in a uplink NOMA system.}.
According to the principle of uplink NOMA, the MEC server firstly decodes  $x_{{\rm{A}}}$ and then subtracts this component from the received signal to decode $x_{{\rm{B}}}$. The achievable capacity from the user ${{\rm{A}}}$ to the MEC server is expressed as}}
\begin{small}
\begin{align}
\tau_{{x_{{{\rm{A}}}}}} = {t_1}{B_c}{\log _2}\left( {1 + \gamma _{{\rm{SINR}}}^{{x_{{{\rm{A}}}}}}} \right).
\end{align}
\end{small}$B_c$ is the bandwidth and $ \gamma _{{\rm{SINR}}}^{{x_{{{\rm{A}}}}}}$ is the received signal to interference and noise ratio (SINR) at the MEC server to decode $x_{{\rm{A}}}$, given by
 \begin{small}
 \begin{align}
\gamma _{{\rm{SINR}}}^{{x_{{{\rm{A}}}}}} = \frac{{\left( {1 + d_{{{\rm{B}}}}^\alpha } \right)\lambda \rho {{\left| {{h_{{{\rm{A}}}}}} \right|}^2}}}{{\left( {1 - \lambda } \right)\left( {1 + d_{{{\rm{A}}}}^\alpha } \right)\rho \left| {{h_{{{\rm{B}}}}}} \right|^2 + \left( {1 + d_{{{\rm{A}}}}^\alpha } \right)\left( {1 + d_{{{\rm{B}}}}^\alpha } \right)}},
 \end{align}
\end{small}where $\rho {\rm{ = }}\frac{P}{{{\sigma ^{\rm{2}}}}}$ is the input signal to noise ratio (SNR).

 If $x_{{\rm{A}}}$ is successfully decoded, i.e., $\tau_{{x_{{{\rm{A}}}}}}\ge\beta_{{\rm{A}}} M_{{\rm{A}}}$, $x_{{\rm{A}}}$ will be subtracted by applying successive interference cancellation (SIC) and the achievable capacity from the user ${{\rm{B}}}$ to the MEC server can be written as
\begin{small}
 \begin{align}
 {\tau_{{x_{{{\rm{B}}}}}}} = {t_1}{B_c}{\log _2}\left( {1 + \gamma _{{\rm{SNR}}}^{{x_{{{\rm{B}}}}}}} \right),
 \end{align}
\end{small}where $\gamma _{{\rm{SNR}}}^{{x_{{{\rm{B}}}}}} = \frac{{\left( {1 - \lambda } \right)\rho \left| {{h_{{{\rm{B}}}}}} \right|^2}}{{ {1 + d_{{{\rm{B}}}}^\alpha } }}$ is the SNR to decode $x_{{\rm{B}}}$.

 If $x_{{\rm{A}}}$ and $x_{{\rm{B}}}$ are decoded successfully at the MEC server, the total  tasks to be executed in the MEC server, denoted by $\tau^{\rm{total}}$, is given as
 \begin{small}
 \begin{align}
\tau^{\rm{total}}=\sum\limits_{k = {\rm{A}},{\rm{B}}} { {{\beta _k}} {M_{{k}}}}.
 \end{align}
 \end{small}
\subsubsection{Task  Executing  Phase}
During the task executing   phase, the local tasks, $(1-\beta_k) M_{{k}}$, and the offloaded tasks, $\tau^{\rm{total}}$, are carried out at the user $k$ and the MEC server, respectively. Therefore, the required time for task execution at the user $k$ and the MEC server can be expressed respectively as:
\begin{small}
\begin{align}\label{6}
t_2^k = \frac{{\left( {1 - \beta_k } \right){M_{{k}}}}C}{{{f_{\rm{user}}}}},\;t_2^{{\rm{MEC}}} = \frac{\tau^{\rm{total}}C}{{{f_{{\rm{MEC}}}}}},
\end{align}
\end{small}where $C$  denotes the number of CPU cycles required for computing one input bit;
${f_{\rm{user}}}$ and ${f_{{\rm{MEC}}}}$ are the CPU frequencies at the users and the MEC server, respectively. Without loss of generality, we assume that ${{f_{{\rm{MEC}}}}}=N{{{f_{\rm{user}}}}}$ with $N>1$ to characterize the difference of their  computational capabilities.
%To meet the  tasks executing demand, we have %the time for tasks executing at the user $k_i$ and the MEC server should less than or equal to $t_2$, i.e,
%\begin{align}
%\max \left\{ {t_2^{\rm{A}},t_2^{\rm{B}},t_2^{{\rm{MEC}}}} \right\} = {t_2}.
%\end{align}
%For ease of  analysis in what follows, we assume the same length of the users' tasks, i.e., $M_{{\rm{A}}}=M_{{\rm{B}}}=M$.
\section{ Successful Computation Probability Maximization}
%In this section, we firstly answer how many tasks should be offloaded to the MEC server to maximize the probability that the users successfully  compute all the tasks $M_{k}$ within $T$. Next, we will study the impact of user selection scheme on the NOMA Based MEC Networks.
We introduce a  successful computation probability   to evaluate the  performance of the considered NOMA based MEC system.  On this basis, we firstly answer how many tasks should be offloaded to the MEC server and how much time should be scheduled for  offloading and  executing to maximize the successful computation probability, then  derive the optimal power allocation ratio for users $\rm{A}$ and $\rm{B}$.  For ease of  analysis, we assume  the same length of the users' tasks  by following the recent works \cite{8492422,8467377}, i.e., $M_{{\rm{A}}}=M_{{\rm{B}}}=M$.

The successful computation probability, denoted by $\mathcal{P}_s$, is  defined as the  probability that  all the tasks $M_{k}$ are successfully  executed within a given time $T$, given by
\begin{align}\label{8}
{{\cal P}_s} \!=\!\Pr \left( \!{{\tau _{{x_{{{\rm{B}}}}}}} \!\ge\! {\beta _{\rm{B}}}M,{\tau _{{x_{{{\rm{A}}}}}}}\! \ge \!{\beta _{\rm{A}}}M,\max \left\{ \! {t_2^{\rm{A}},t_2^{\rm{B}},t_2^{{\rm{MEC}}}} \right\} \le {t_2}} \!\right).
\end{align}

\textbf{Proposition 1.} For any given parameters, i.e., $t_1$, $t_2$, ${\beta _{\rm{A}}}$, ${\beta _{\rm{B}}}$ and $\lambda$, the closed-form expression of the successful computation probability can be written as
\begin{align}\label{9}
{{\cal{P}}_s} \!=\! \left\{ {\begin{array}{*{20}{c}}
\begin{array}{l}
\!\!\!\!\! \exp \left( { - \frac{{\left( {1 + d_{{{\rm{B}}}}^\alpha } \right){\gamma _2}}}{{\left( {1 - \lambda } \right)\rho }} - \frac{{{\gamma _1}\left( {1 + d_{{{\rm{A}}}}^\alpha } \right)\left( {{\rm{1 + }}{\gamma _2}} \right)}}{{\lambda \rho }}} \right) \times \\
\!\!\!\!\!\frac{{\left( {1 + d_{{{\rm{B}}}}^\alpha } \right)\lambda }}{{\left( {1 + d_{{{\rm{B}}}}^\alpha } \right)\lambda  + {\gamma _1}\left( {1 - \lambda } \right)\left( {1 + d_{{{\rm{A}}}}^\alpha } \right)}},\;{\rm{if}}\;\max \left\{ {t_2^{\rm{A}},t_2^{\rm{B}},t_2^{{\rm{MEC}}}} \right\} \le {t_2},\;
\end{array}\\
{\!\!\!\!\! \! \! 0,\;\;\;\;\;\;\;\;\;\;\;\;\;\;\;\;\;\;\;\;\;\;\;\;\;\;\;\;\;\;\;\;\;\;\;\;\;\;\;\;\;\;\;\;\;\;\;\;\;\;\;\;\;\;\;\;\;\;\;\;\;\;{\rm{otherwise}}},
\end{array}} \right.
\end{align}
where ${\gamma _1} = {2^{\frac{{{\beta _{\rm{A}}}M}}{{{t_1}{B_c}}}}} - 1$ and  ${\gamma _2} = {2^{\frac{{{\beta _{\rm{B}}}M}}{{{B_c}{t_1}}}}} - 1$.

\emph{Proof.} Please see Appendix A. \hfill {$\blacksquare $}

\emph{Remark 1.} Proposition 1 serves the following  purposes. Firstly, we provide a closed-form expression to characterize the successful transmission probability of uplink NOMA based MEC networks.
%The second purpose is to obtain insightful understandings on our considered network, as shown in Corollary 1.
Besides, the closed-form expression in Proposition 1 offers a  possibility to obtain the optimal parameters in terms of the maximum successful transmission probability, which  are particularly helpful for designing our considered NOMA based MEC network. It is worth noting that, different from the existing works  \cite{8434285, 8537962, 8267072,8492422} with a focus on the design of NOMA based MEC networks based on the instantaneous CSI, our designed network is  based
on the statistic CSI  and  removes the need to know the  accurate instantaneous CSI,
 alleviating the burdens of signallings.

\emph{Corollary 1.}  If  $\frac{{MC}}{{{f_{{\rm{user}}}}}} \le T$  holds, we have ${{\cal{P}}_s}=1$. In this case, the users execute all the tasks  locally within the maximum supportable latency $T$ and the desirable working mode   is the complete local computation in terms of successful computation probability. Therefore,  the optimal network parameters are as follows: $\beta^*_{\rm{A}}=\beta^*_{\rm{B}}=0$, $ t_1^*=0$, $ t_2^* = \frac{{{M_{{k}}}}C}{{{f_{\rm{user}}}}}$, and $\lambda^*$ can be taken at any value.

In order to provide more  insights in designing  our considered network, we formulate an problem to maximize the successful computation probability by jointly optimizing the parameters of the proposed offloading parameters, i.e.,  $t_1$, $t_2$, ${\beta _{\rm{A}}}$, ${\beta _{\rm{B}}}$ and $\lambda$, for the case\footnote{As the optimal parameters of the proposed scheme have been obtained in Corollary 1 when $\frac{{MC}}{{{f_{{\rm{user}}}}}} \le T$, here we only focus on the case with  $\frac{{MC}}{{{f_{{\rm{user}}}}}} > T$. } with $\frac{{MC}}{{{f_{{\rm{user}}}}}} > T$ in the following.
\begin{small}
\begin{align}
\begin{array}{*{20}{l}}
{{{\bf{P}}_{\bf{0}}}:\;\;\mathop {\max }\limits_{{t_1},{t_2},{\beta _{\rm{A}}},{\beta _{\rm{B}}},\lambda } \;{{\cal{P}}_s}}\\
{\;\;\;\;\;\;\;\;\;\;\;\;\;\;\;\;{t_1} + {t_2} \le T\;},\\
\;\;\;\;\;\;\;\;\;\;\;\;\;\;\;\frac{{MC}}{{{f_{{\rm{user}}}}}} > T\;{\rm{0}} < \lambda  < {\rm{1}},\;\\
{\;\;\;\;\;\;\;\;\;\;\;\;\;\;\;\;0 < t_1,t_2,{\beta _{\rm{A}}},{\beta _{\rm{B}}} \le 1},\\
{\;\;\;\;\;\;\;\;\;\;\;\;\;\;\;\;\max \left\{ {t_2^{\rm{A}},t_2^{\rm{B}},t_2^{{\rm{MEC}}}} \right\} \le {t_2}.\;\;\;\;\;\;\;\;\;\;\;\;\;}
\end{array}
\end{align}
\end{small}It is not hard to find that  the problem ${{\bf{P}}_{\bf{0}}}$ is a non-convex problem due to the non-convex objective function. In general, there is no
standard algorithm to solve non-convex optimization problems efficiently.  We provide the following two propositions to obtain the
optimal solutions as follows.

\textbf{Proposition 2.} The equality  $t_1+t_2=T$ holds for the successful computation probability maximization.

\emph{Proof.} Please see Appendix B. \hfill {$\blacksquare $}

\textbf{Proposition 3.} The maximum  successful computation probability can be always achieved by satisfying the following  equality, i.e.,  ${t_2^{\rm{A}}=t_2^{\rm{B}}=t_2^{{\rm{MEC}}}} = {t_2}$.

\emph{Proof.} Please see Appendix C. \hfill {$\blacksquare $}

\emph{Remark 2.} The intuition behind Proposition 3 is that the two users and the MEC server complete the tasks executing within the same time in order to make the best use of the tasks executing time $t_2$, reducing the length of the tasks needed to be offloaded. This is beneficial  to enhance the successful computation probability, as shown in Lemma 1 (Please find Lemma 1 in the fourth paragraph of Appendix C).

Based on propositions 2 and 3, we can derive the optimal  offloading time $t_1$, the task executing time $t_2$,  $\beta_{{\rm{A}}}$ and $\beta_{\rm{B}}$, as summarized in Theorem 1.

\textbf{Theorem 1.} To maximize the successful computation probability, we have
\begin{small}
\begin{align}
\left\{ {\begin{array}{*{20}{c}}
{\begin{array}{*{20}{c}}
{\!\!\!\!\!\!t_{\rm{1}}^{\rm{*}} = T - \frac{{2MC}}{{{f_{{\rm{user}}}}\left( {N{\rm{ + 2}}} \right)}}},\\
{\!\!\!\!\!\!t_2^{\rm{*}} = \frac{{2MC}}{{{f_{{\rm{user}}}}\left( {N{\rm{ + 2}}} \right)}},\;\;\;\;\;}
\end{array}}\\
{\!\!\!\!\!\!\beta _{\rm{A}}^* = \beta _{\rm{B}}^* = \frac{N}{{2 + N}},\;\;\;}
\end{array}} \right.
\end{align}
\end{small}where $*$ denotes the optimal solution corresponding to the optimization variables.
\begin{figure*}
\begin{minipage}[t]{0.332\linewidth}
\centering
\includegraphics[width=2.33in,height=1.85in]{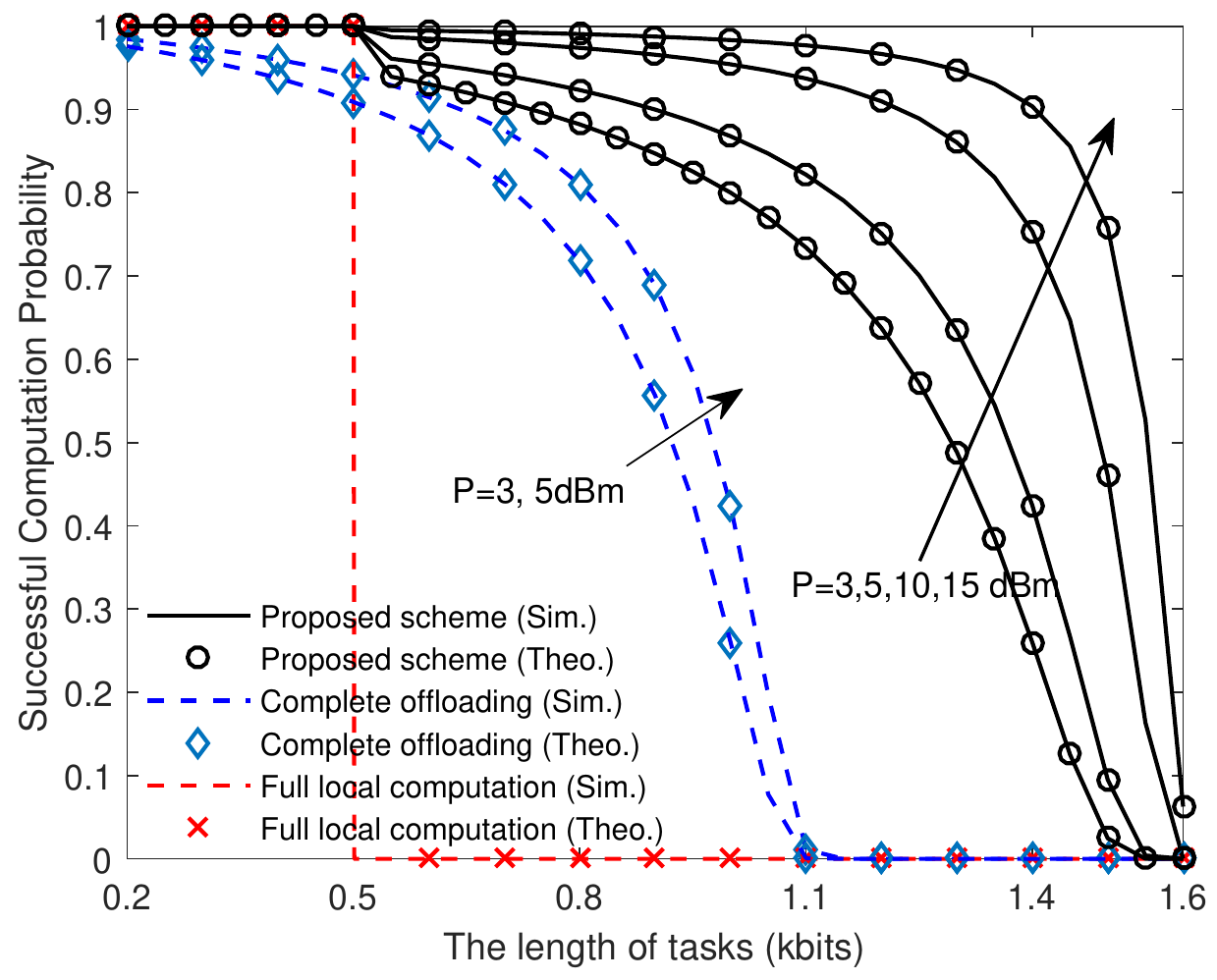}
\centering \caption{${\cal{P}}_s$ versus the length of tasks.}
\end{minipage}
\begin{minipage}[t]{0.332\linewidth}
\centering
\includegraphics[width=2.33in,height=1.85in]{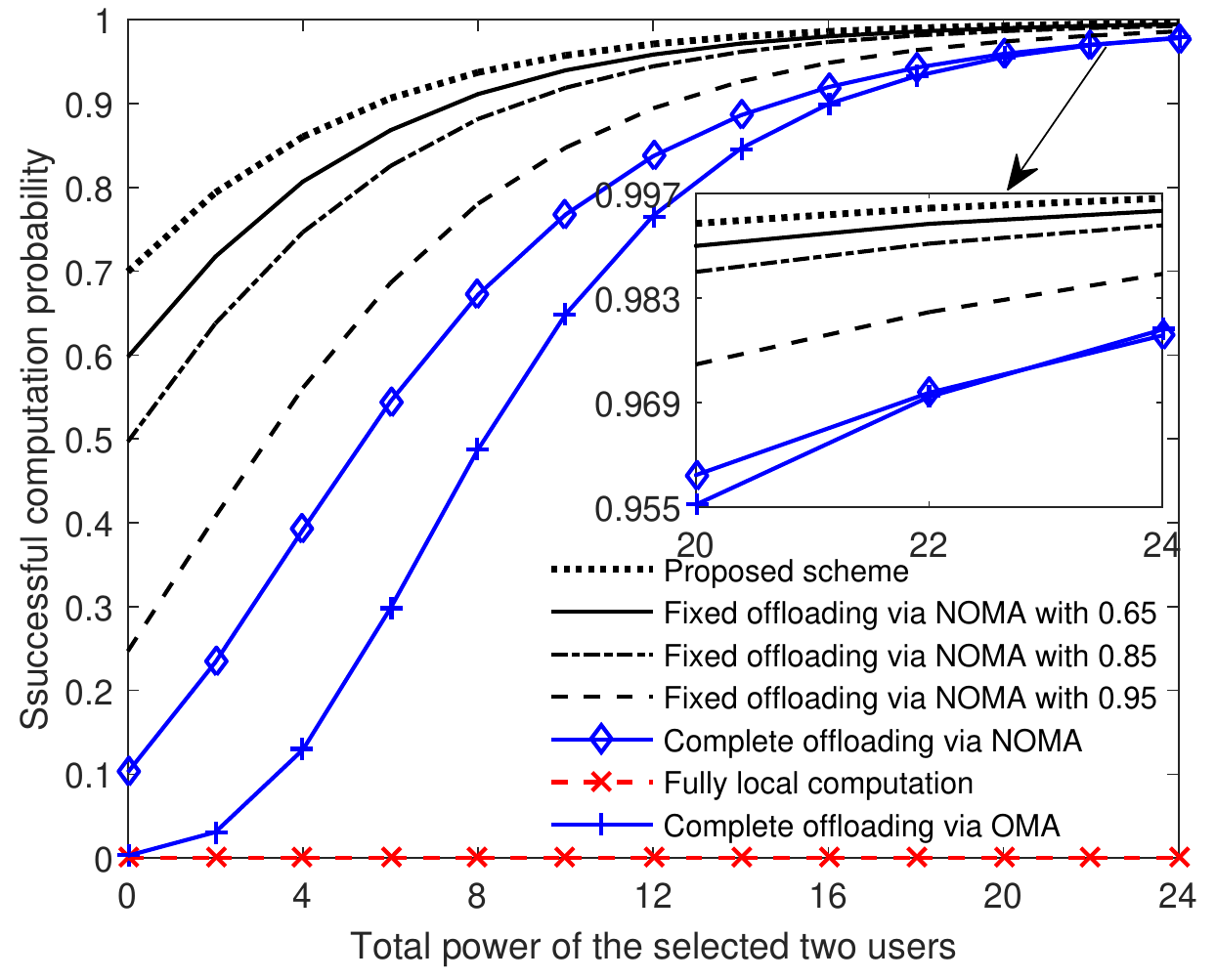}
\centering \caption{ ${\cal{P}}_s$ versus total power of the users.}
\end{minipage}
\begin{minipage}[t]{0.332\linewidth}
\centering
\includegraphics[width=2.33in,height=1.85in]{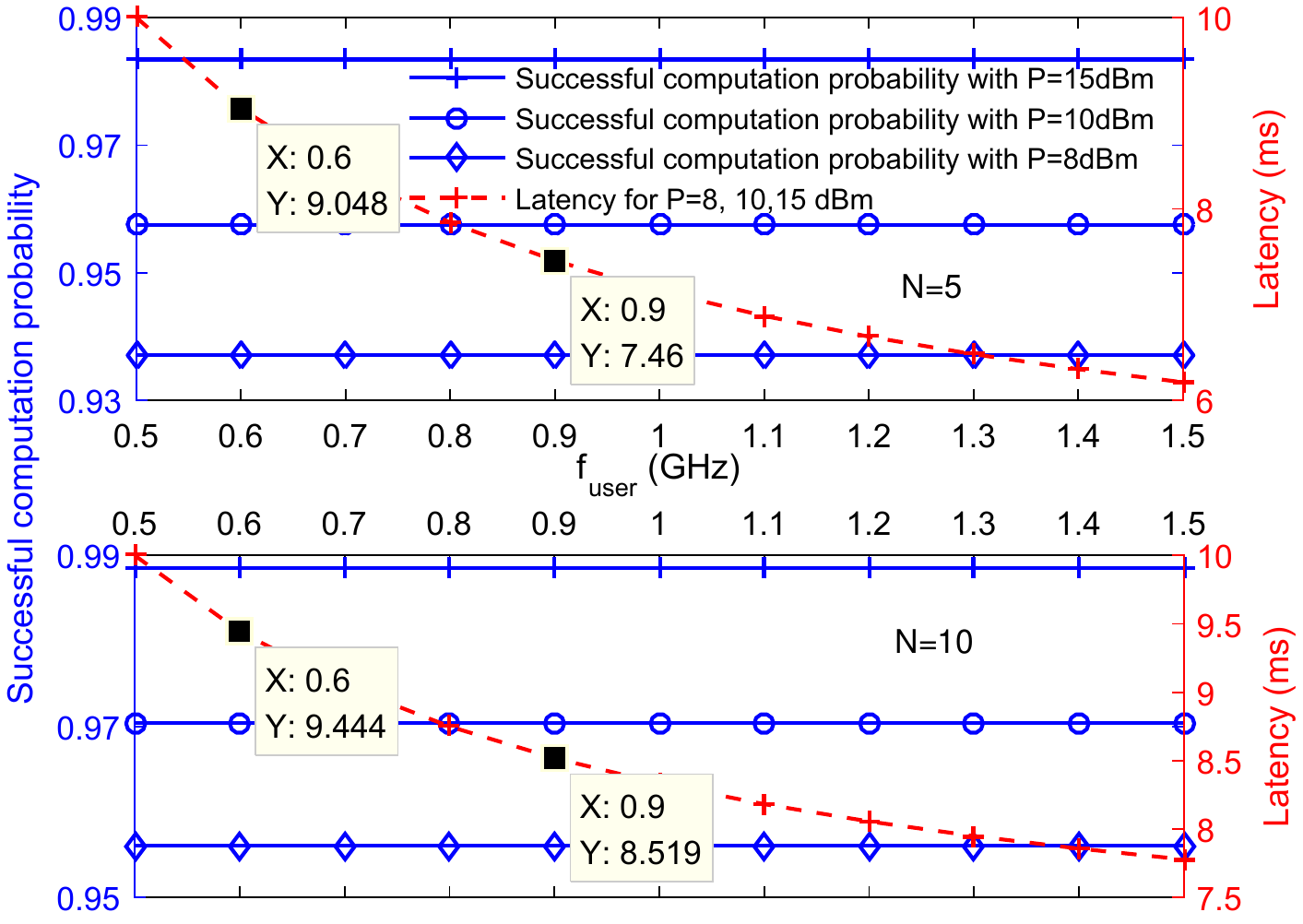}
\centering \caption{The way to reduce latency.}
\end{minipage}
\end{figure*}

\emph{Proof.}  When $t_2^{\rm{A}*}= t_2^{\rm{B}*} = t_2^{{\rm{MEC}*}}=t_2^*$ is satisfied, we have ${t_2^ *  = \frac{{2MC}}{{{f_{{\rm{user}}}}\left( {N{\rm{ + 2}}} \right)}}}$ and $\beta _{\rm{A}}^ *  = \beta _{\rm{B}}^* =1 - \frac{{{f_{{\rm{user}}}}t_2^ * }}{{MC}}$. Therefore, we have ${\beta _{\rm{A}}^* = \beta _{\rm{B}}^* = \frac{N}{{2 + N}}\;}$. Combing Theorem 1 and  Proposition 2, $t_1^*$ can also be determined and the proof is complete. \hfill {$\blacksquare $}

\emph{Remark 3.} Theorem 1 reveals the following  facts. Firstly, the optimal parameters, i.e., $t_1^*$, $t_2^*$, $\beta _{\rm{A}}^*$, $\beta _{\rm{B}}^*$, are independent of the locations of the users and the power allocation coefficient $\lambda$. This means that the derived  results in Theorem 1 can be directly used for randomly deployed users, and that $\lambda^*$ can be obtained by solving $\frac{{\partial {{\cal{P}}_s}}}{{\partial \lambda }}=0$ if $\lambda^*$ exists.
Secondly, $N$ determines how many tasks needed to be offloaded for the two users. In particular, the users are willing to offload more tasks to improve the successful computation probability as $N$ increases. When $N \to \infty$, we have  ${\beta _{\rm{A}}^* = \beta _{\rm{B}}^*}=1$ and the complete offloading is desirable, while the partial offloading is better than complete offloading when $N$ is finite. This means that the complete offloading  is not an optimal working mode if the users have  strong computational capability.
%In other words, the partial offloading is the optimal scheme instead of complete offloading in  practice since $N$ is finite.
 Thirdly, for a given $T$, we have derived an optimal time allocation scheme to achieve  the tradeoff between the  offloading time $t_1$ and the tasks executing time $t_2$ in terms of successful computation probability. Lastly, for a given $M$,  the latency can be reduced without decreasing the successful computation probability when we increase  $f_{\rm{user}}$ and $f_{\rm{MEC}}$ while $N$ remains unchanged. The reason is as follows. When $f_{\rm{user}}$ increases, $\beta _{\rm{A}}^*$, $\beta _{\rm{B}}^*$  and the length of tasks to be offloaded remain unchanged, while the tasks executing time $t_2$ decreases. In this case,   the successful computation probability remains unchange if   the  offloading time $t_1$ remains unchanged. Therefore, $t_1$+$t_2$ decreases and the latency is reduced.

Let us turn our attention to study how the MEC server allocates the power for the two fixed users   to maximize the successful computation probability.

\textbf{Theorem 2.}  There  is a unique optimal power allocation coefficient $\lambda^* $ for ${\bf{P}}_0$ and its closed-form expression is written as
\begin{small}
\begin{align}\label{13}
{\lambda ^*} =
\left\{\!\!\! {\begin{array}{*{20}{c}}\!
{{x_N} + \sqrt[3]{{\frac{{ - {y_N} + \sqrt {y_N^2 - {g^2}} }}{{2{m_1}}}}} + \sqrt[3]{{\frac{{ - {y_N} - \sqrt {y_N^2 - {g^2}} }}{{2{m_1}}}}},\;{\rm{if}}\;y_N^2 > {g^2}}\\
\!\!\!\!\!\!\!{\left\{ {0 < \lambda  < 1\left| {\lambda  = {x_N} + \delta \;{\rm{or}}\;{x_N}   - 2\delta } \right.} \right\},\;{\rm{if}}\;y_N^2 = {g^2}}\\
\!\left\{ {0 < \lambda  < 1\left| {\lambda  = {x_N} + 2\delta \cos \left( {\phi  - \frac{{2\pi i}}{3}} \right),\;i = 0,1,2} \right.} \right\},\\{\rm{otherwise}}
\end{array}} \right.
\end{align}
\end{small}where ${a_1} = \frac{{{\gamma _1}\left( {1 + d_{{{\rm{A}}}}^\alpha } \right)}}{{1 + d_{{{\rm{B}}}}^\alpha }}$, $a_2=\frac{{\left( {1 + d_{{{\rm{B}}}}^\alpha } \right){\gamma _2}}}{\rho }$, ${a_3} = \frac{{{\gamma _1}\left( {1 + d_{{{\rm{A}}}}^\alpha } \right)\left( {{\rm{1 + }}{\gamma _2}} \right)}}{\rho }$,
${m_1} = {a_1} + {a_3} -{a_2} + {a_1}{a_2} - {a_1}{a_3}$, ${m_2} = 3{a_1}{a_3} - 2{a_1} - {a_1}{a_2} - 2{a_3}$, ${m_3} = {a_1} - 3{a_1}{a_3} + {a_3}$, ${\theta ^{\rm{2}}}{\rm{ = 3}}{\delta ^{\rm{2}}}$, $g =   2{m_1}{\delta ^3}$, ${\delta ^2} = \frac{{m_2^2 - 3{m_1}{m_3}}}{{9m_1^2}}$, ${x_N} =  - \frac{{{m_2}}}{{3{m_1}}}$, ${y_N} = \frac{{2m_2^{\rm{3}}}}{{27m_1^2}} - \frac{{{m_2}{m_3}}}{{3{m_1}}} + {a_1}{a_3}$, $\phi  = \frac{1}{3}\arccos \left( { - \frac{{{y_N}}}{g}} \right)$ and the sign of $\delta$ is the same of the sign of $\sqrt[{\rm{3}}]{{\frac{{{y_N}}}{{2{m_1}}}}}$.

\emph{Proof.} Please see Appendix D. \hfill {$\blacksquare $}

Based on the above analysis, the optimal network parameters can be determined and summarized as
\begin{small}
\begin{align}\label{14}\notag
&\left( {\beta _{\rm{A}}^*,\beta _{\rm{B}}^*,t_{\rm{1}}^{\rm{*}},{t_2^{k}}^*, {\lambda ^*}} \right)\\
&  =\left\{ {\begin{array}{*{20}{c}}
{\left( {0,0,0,\frac{{{M}}C}{{{f_{\rm{user}}}}},{\rm{any}}\;{\rm{value}}} \right),\;\;\;\;\;\;\;\;\;\;\;\;\;\;\;{\rm{if}}\;\frac{{MC}}{{{f_{{\rm{user}}}}}} \le T},\\
{{\rm{shown}}\;{\rm{in}}\;{\rm{Theorem1}}\;{\rm{and}}\;{\rm{Theorem2}},{\rm{if }}\frac{{MC}}{{{f_{{\rm{user}}}}}} > T}.
\end{array}} \right.
\end{align}
\end{small}Substituting \eqref{14} into \eqref{9}, the maximum successful computation probability is given as
\begin{small}
\begin{align}\label{15}\notag
  {\cal{P}}_s^*&= \exp \left( { - \frac{{\left( {1 + d_{{{\rm{B}}}}^\alpha } \right){\gamma ^*}}}{{\left( {1 - \lambda } \right)\rho }} - \frac{{{\gamma ^*}\left( {1 + d_{{{\rm{A}}}}^\alpha } \right)\left( {{\rm{1 + }}{\gamma _2}} \right)}}{{\lambda \rho }}} \right)\\
 & \times \frac{{\left( {1 + d_{{{\rm{B}}}}^\alpha } \right){\lambda ^*}}}{{\left( {1 + d_{{{\rm{B}}}}^\alpha } \right){\lambda ^*} + {\gamma ^*}\left( {1 - {\lambda ^*}} \right)\left( {1 + d_{{{\rm{A}}}}^\alpha } \right)}},
\end{align}
\end{small}where ${\gamma ^*} = \left\{ {\begin{array}{*{20}{c}}
{{2^{\frac{{NM}}{{\left( {T\left( {2 + N} \right) - \frac{{2MC}}{{{f_{{\rm{user}}}}}}} \right){B_c}}}}} - 1,{\rm{if}}\;\frac{{MC}}{{{f_{{\rm{user}}}}}} > T},\\
{0,\;\;\;\;\;\;\;\;\;\;\;\;\;\;\;\;\;\;\;\;\;\;\;\;\;\;\;\;\;\;\;{\rm{otherwise}}}.
\end{array}} \right.$

\emph{Corollary 2.} If $\rho \to \infty $, the maximum successful computation probability approaches one. This means our considered NOMA based MEC network with optimal parameters is able to meet any required successful computation probability by adjusting the transmit power.

\emph{Proof.} If  $\rho \to \infty $, $a_2$ and $a_3$ approach zero and (D.3) can be approximated as
${a_1}{\lambda ^3} -2 {a_1}{\lambda ^2} + {a_1}\lambda  \approx 0$. In this case,  $\lambda^*\to 1$ holds. Combing \eqref{15}, we have $\mathop {\lim }\limits_{\rho  \to \infty } {{\cal{P}}_s^*} = 1$.  \hfill {$\blacksquare $}
% \begin{figure}[t]
%    \centering
%  \includegraphics[width=0.3\textwidth]{fig1.pdf}\\
%  \caption{${\cal{P}}_s$ versus the length of tasks.}
%\end{figure}
\section{Simulation Results}
%The simulation parameters for Fig. 2 are as follows: $\lambda=0.3$, $N=5$, $\beta=\beta^*$, $T=10$ms, $t_1=t_1^*$, $t_2=t_2^*$, $f_{\rm{user}}=0.5$GHz, $C=1000$cycle/bit, ${\sigma ^2}=10^{-9}$W, $d_{{\rm{A}}}=5$m, $d_{{\rm{B}}}=25$m, $\alpha=4$.
%The simulation parameters for Fig. 3 are as follows:  $T=10$ms,   $C=1000$cycle/bit, ${\sigma ^2}=10^{-9}$W, $d_{{\rm{A}}}=5$m, $d_{{\rm{B}}}=25$m, $\alpha=4$, $M=10$kbits, $t_1=t_1^*$, $t_2=t_2^*$, $\lambda=\lambda^*$, $N=5$, $\beta=\beta^*$.
%The simulation parameters for Fig. 4 are as follows: $\lambda=\lambda^*$, $\beta=\beta^*$,  $t_1=t_1^*$, $t_2=t_2^*$, $f_{\rm{user}}=0.5$GHz, $C=1000$cycle/bit, ${\sigma ^2}=10^{-9}$W, $d_{{\rm{A}}}=5$m, $d_{{\rm{B}}}=25$m, $\alpha=4$, $M=10$kbits.
In this section, simulation  results are provided to verify our derived results and investigate the successful computation probability  the proposed offloading scheme can achieve. Unless otherwise specified, the following parameters are used throughout the simulation: $N=5$, $T=10$ms, $f_{\rm{user}}=0.5$GHz, $C=1000$cycle/bit, $M=10$kbits, ${\sigma ^2}=10^{-9}$W, $d_{{\rm{A}}}=5$m, $d_{{\rm{B}}}=25$m, $\alpha=4$,  $t_1=t_1^*$, $t_2=t_2^*$, $\lambda=\lambda^*$ and $\beta_{\rm{A}}=\beta_{\rm{B}}=\beta^*$.

% \begin{figure}[t]
%    \centering
%  \includegraphics[width=0.3\textwidth]{fig2.pdf}\\
%  \caption{${\cal{P}}_s$ versus total power of the users}
%\end{figure}
Fig. 2 shows the successful computation probability versus the length of tasks, where three schemes are considered: (i) the proposed scheme, (ii) the full local computation scheme, (iii) the complete offloading scheme. The power allocation $\lambda$ is set as 0.3. It can be observed that the successful computation probability $\mathcal{P}_s$ decreases with the increase of  the length  $M$ of  tasks.  This is because the offloading time $t_1$ also decreases with the increase of $M$, resulting in the decrease of $\mathcal{P}_s$. It can also be seen that the proposed scheme achieves the highest successful computation probability than the complete local computation scheme and the complete offloading scheme. This is because the proposed scheme can decide how many tasks should be offloaded by considering  the length of tasks, the difference of their computational capabilities, etc. For example, when $M$ is small, the tasks tend to be locally executed, while when $M$ is large enough, the tasks tend to be completely offloaded.

Fig. 3 shows the successful computation probability versus the total power $P$ under five schemes: (1) the proposed scheme, (2) the fixed offloading via NOMA scheme, (3) the complete offloading via NOMA scheme, (4) the complete offloading via OMA scheme, (5) the full local computation scheme. For the fixed offloading via NOMA scheme, the offloading ratio is fixed as $0.65$, $0.85$ and $0.95$, respectively. One observation is that the successful computation probability increases with the increase of $P$.  This is due to the fact that a larger $P$ brings a larger $\rho$, resulting in a larger $\mathcal{P}_s$.
Among these schemes, the proposed scheme achieves the highest successful computation probability.% One can also see that, for the complete offloading, the successful computation probability achieved by the uplink  NOMA is  larger than that of the OMA.

Fig. 4 illustrates the effectiveness of the proposed scheme to reduce latency in \emph{Remark 3} and shows the successful computation probability $\mathcal{P}_s$ (the left vertical axis) and the latency (the right vertical axis) trend with the increase of $f_{\rm{user}}$. The offloading time is set as $t_1=10- \frac{{2MC}}{{5\times 10^8\left( {N{\rm{ + 2}}} \right)}}$ ms. It can be observed that with the increase of $f_{\rm{user}}$, $\mathcal{P}_s$ remains unchanged while the latency decreases. This is because both the offloading ratios, $\beta_{\rm{A}}^*$ and $\beta_{\rm{B}}^*$, as well as  the offloading time $t_1$ keep unchanged while the tasks executing time $t_2$ decreases with the increase of $f_{\rm{user}}$.
It can be also seen that for the same set of $f_{\rm{user}}$, with the increase of $N$, $\mathcal{P}_s$ increases with the cost of higher latency. This is because with a larger $N$, the users tend to offload more tasks to achieve a higher $\mathcal{P}_s$, leading to the increase of the offloading time $t_1$.
% \begin{figure}[t]
%    \centering
%  \includegraphics[width=0.34\textwidth]{fig3.pdf}\\
%  \caption{Illustration of the way to reduce latency in Remark 3}
%\end{figure}

\section{Conclusions}
We have proposed a new offloading scheme for a NOMA based MEC network, which can operate in the partial computation offloading, the full local computation or the complete offloading. We have derived the successful computation probability for the proposed offloading scheme. We also formulated an optimization problem to maximize the successful computation probability by jointly optimizing the parameters of the proposed offloading scheme and obtained the optimal solutions in closed forms. Simulation results were presented to show that our proposed scheme outperforms the existing schemes in terms of successful computation probability.

\appendices
\begin{appendices}
\section{Proof of the Proposition 1}
According to the Law of total probability, we derive the successful computation probability from  the following two cases, i.e.,  Case I: ${\max \left\{ {t_2^{\rm{A}},t_2^{\rm{B}},t_2^{{\rm{MEC}}}} \right\} > {t_2}}$ and  Case II: ${\max \left\{ {t_2^{\rm{A}},t_2^{\rm{B}},t_2^{{\rm{MEC}}}} \right\} \le {t_2}}$. Since there are no variables, i.e., ${{{\left| {{h_{{{\rm{A}}}}}} \right|}^2}}$ and ${{{\left| {{h_{{{\rm{B}}}}}} \right|}^2}}$, involved in the following two inequalities: ${\max \left\{ {t_2^{\rm{A}},t_2^{\rm{B}},t_2^{{\rm{MEC}}}} \right\} > {t_2}}$ and  ${\max \left\{ {t_2^{\rm{A}},t_2^{\rm{B}},t_2^{{\rm{MEC}}}} \right\} \le {t_2}}$, we have
\begin{small}
\begin{align}\left\{ {\begin{array}{*{20}{c}}\tag{A.1}
{{\rm{Case}}\;{\rm{I}}:\;\;\Pr \left( {\max \left\{ {t_2^{\rm{A}},t_2^{\rm{B}},t_2^{{\rm{MEC}}}} \right\} > {t_2}} \right) = 1},\\
{{\rm{Case\;II}}:\Pr \left( {\max \left\{ {t_2^{\rm{A}},t_2^{\rm{B}},t_2^{{\rm{MEC}}}} \right\} \le {t_2}} \right) = 1}.
\end{array}} \right.
\end{align}
\end{small}Applying the Law of total probability and considering (A.1), it is not hard to find that the successful computation probability of  Case I equals zero.
By the same argument, the successful computation probability of Case II can be rewritten as (A.2), as shown at the top of the next page.
\begin{figure*}[t]
\normalsize
\begin{small}
\begin{align} \notag
{{\cal P}_s}&\! =\!\Pr \left( {\max \left\{\! {t_2^{\rm{A}},t_2^{\rm{B}},t_2^{{\rm{MEC}}}}\! \right\} \le {t_2}}\! \right)\Pr \left( \!{\left. {\frac{{\left( {1 + d_{{{\rm{B}}}}^\alpha } \right){\gamma _2}}}{{\left( {1 - \lambda } \right)\rho }} \le {{\left| {{h_{{{\rm{B}}}}}}\! \right|}^2} \le \frac{{\left( {1 + d_{{{\rm{B}}}}^\alpha } \right)\lambda \rho {{\left| {{h_{{{\rm{A}}}}}} \right|}^2} - {\gamma _1}\left( {1 + d_{{{\rm{A}}}}^\alpha } \right)\left( {1 + d_{{{\rm{B}}}}^\alpha } \right)}}{{{\gamma _1}\left( {1 - \lambda } \right)\left( {1 + d_{{{\rm{A}}}}^\alpha } \right)\rho }}} \right|\max \left\{ {t_2^{\rm{A}},t_2^{\rm{B}},t_2^{{\rm{MEC}}}} \right\} \le {t_2}} \right) \\ \notag
%&=\int_{\frac{{{\gamma _1}\left( {1 + d_{{{\rm{A}}}}^\alpha } \right)\left( {{\rm{1 + }}{\gamma _2}} \right)}}{{\lambda \rho }}}^\infty  {\int_{\frac{{\left( {1 + d_{{{\rm{B}}}}^\alpha } \right){\gamma _2}}}{{\left( {1 - \lambda } \right)\rho }}}^{\frac{{\left( {1 + d_{{{\rm{B}}}}^\alpha } \right)\lambda \rho x - {\gamma _1}\left( {1 + d_{{{\rm{A}}}}^\alpha } \right)\left( {1 + d_{{{\rm{B}}}}^\alpha } \right)}}{{{\gamma _1}\left( {1 - \lambda } \right)\left( {1 + d_{{{\rm{A}}}}^\alpha } \right)\rho }}} {\exp \left( { - t} \right)dt} } \exp \left( { - x} \right)dx\\ \notag
& =\int_{\frac{{{\gamma _1}\left( {1 + d_{{{\rm{A}}}}^\alpha } \right)\left( {{\rm{1 + }}{\gamma _2}} \right)}}{{\lambda \rho }}}^\infty  {{\rm{exp}}\left( { - \frac{{\left( {1 + d_{{{\rm{B}}}}^\alpha } \right){\gamma _2}}}{{\left( {1 - \lambda } \right)\rho }} - x} \right) \!-\! {\rm{exp}}\left( { \!- \frac{{\left( {1 + d_{{{\rm{B}}}}^\alpha } \right)\lambda \rho x - {\gamma _1}\left( {1 + d_{{{\rm{A}}}}^\alpha } \right)\left( {1 + d_{{{\rm{B}}}}^\alpha } \right)}}{{{\gamma _1}\left( {1 - \lambda } \right)\left( {1 + d_{{{\rm{A}}}}^\alpha } \right)\rho }} - x} \right)} dx\\ \notag
&=\exp \left( { - \frac{{\left( {1 + d_{{{\rm{B}}}}^\alpha } \right){\gamma _2}}}{{\left( {1 - \lambda } \right)\rho }} - {\frac{{{\gamma _1}\left( {1 + d_{{{\rm{A}}}}^\alpha } \right)\left( {{\rm{1 + }}{\gamma _2}} \right)}}{{\lambda \rho }}}} \right)- \exp \left( {\frac{{1 + d_{{{\rm{B}}}}^\alpha }}{{\left( {1 - \lambda } \right)\rho }}} \right)\int_{{\frac{{{\gamma _1}\left( {1 + d_{{{\rm{A}}}}^\alpha } \right)\left( {{\rm{1 + }}{\gamma _2}} \right)}}{{\lambda \rho }}}}^\infty  {\exp \left( { - \left( {\frac{{\left( {1 + d_{{{\rm{B}}}}^\alpha } \right)\lambda }}{{{\gamma _1}\left( {1 - \lambda } \right)\left( {1 + d_{{{\rm{A}}}}^\alpha } \right)}} + 1} \right)x} \right)dx} \\
&={\frac{{\left( {1 + d_{{{\rm{B}}}}^\alpha } \right)\lambda }}{{\left( {1 + d_{{{\rm{B}}}}^\alpha } \right)\lambda  + {\gamma _1}\left( {1 - \lambda } \right)\left( {1 + d_{{{\rm{A}}}}^\alpha } \right)}}\exp \left( { - \frac{{\left( {1 + d_{{{\rm{B}}}}^\alpha } \right){\gamma _2}}}{{\left( {1 - \lambda } \right)\rho }} - \frac{{{\gamma _1}\left( {1 + d_{{{\rm{A}}}}^\alpha } \right)\left( {{\rm{1 + }}{\gamma _2}} \right)}}{{\lambda \rho }}} \right)}.\tag{A.2}
\end{align}
\end{small}
\hrulefill
\end{figure*}

\section{Proof of the Proposition 2}
 Proposition 2 is proven by using contradiction. Let $\left( {t_1^ * ,{t_2^ *},{\beta^ * _{\rm{A}}},{\beta^ * _{\rm{B}}},\lambda^ * } \right)$ denote the optimal solution for ${{\bf{P}}_{\bf{0}}}$ and the optimal solution satisfies $t_1^ * +{t_2^ *}<T$.    In addition to the optimal solution, we construct another feasible solution, denoted by
  $\left( {t_1^ + ,{t_2^ +},{\beta ^ +_{\rm{A}}},{\beta^ + _{\rm{B}}},\lambda^ + } \right)$, for ${{\bf{P}}_{\bf{0}}}$, where we have $t_2^ *=t_2^ +$, ${\beta^ * _{\rm{A}}}={\beta^ + _{\rm{A}}}$, ${\beta ^ *_{\rm{B}}}={\beta^ + _{\rm{B}}}$, $\lambda^ *=\lambda^ +$, $t_1^ +>t_1^*$ and $t_1^ + + t_2^ +=T$.  Also, the corresponding successful computation probabilities with $\left( {t_1^ * ,{t_2^ *},{\beta^ * _{\rm{A}}},{\beta^ * _{\rm{B}}},\lambda^ * } \right)$ and $\left( {t_1^ + ,{t_2^ +},{\beta^ + _{\rm{A}}},{\beta ^ +_{\rm{B}}},\lambda^ + } \right)$ are denoted as ${\cal{P}}_s^*$ and ${\cal{P}}_s^+$, respectively. Obviously, the solution  with $\left( {t_1^ + ,{t_2^ +},{\beta^ + _{\rm{A}}},{\beta ^ +_{\rm{B}}},\lambda^ + } \right)$ satisfies all the constraints of  ${{\bf{P}}_{\bf{0}}}$.

  Next, we  discuss the relationship between  ${{\cal{P}}_s}$  and $t_1$ when other parameters are fixed. The first-order derivative of ${{\cal{P}}_s}$ with respect to $t_1$ is calculated as
  \begin{small}
  \begin{align}
 \frac{{\partial {{\cal{P}}_s}}}{{\partial {t_1}}} = \frac{{\partial {{\cal{P}}_s}}}{{\partial {\gamma _1}}} \times \frac{{\partial {\gamma _1}_s}}{{\partial {t_1}}} = \frac{{\partial {{\cal{P}}_s}}}{{\partial {\gamma _1}}} \times \left( { - \frac{{{\beta _{\rm{A}}}M}}{{t_1^2{B_c}}}{2^{\frac{{{\beta _{\rm{A}}}M}}{{{t_1}{B_c}}}}}} \right)\tag{B.1}.
  \end{align}
\end{small}Thus, the  monotonicity of ${{\cal{P}}_s}$ with respect to $t_1$ depends on $ \frac{{\partial {{\cal{P}}_s}}}{{\partial {\gamma _1}}}$, given by
\begin{small}
\begin{align}
\frac{{\partial {{\cal{P}}_s}}}{{\partial {\gamma_1}}} \!=\! - \!\left( {\frac{{{k_1}}}{{{{\left( {1 + {k_1}{\gamma _1}} \right)}^2}}}\! +\! \frac{{{k_3}}}{{1 + {\gamma _1}{k_1}}}} \right)\exp \left( { - {k_2} - {k_3}{\gamma _1}} \right), \tag{B.2}
\end{align}
\end{small}where $k_1={\frac{{\left( {1 - \lambda } \right)\left( {1 + d_{{{\rm{A}}}}^\alpha } \right)}}{{\left( {1 + d_{{{\rm{B}}}}^\alpha } \right)\lambda }}}$, $k_2={\frac{{\left( {1 + d_{{{\rm{B}}}}^\alpha } \right){\gamma _2}}}{{\left( {1 - \lambda } \right)\rho }}}$, and $k_3={\frac{{\left( {1 + d_{{{\rm{A}}}}^\alpha } \right)\left( {{\rm{1 + }}{\gamma _2}} \right)}}{{\lambda \rho }}}$. Obviously, $\frac{{\partial {{\cal{P}}_s}}}{{\partial {t_1}}}>0$ and  ${{\cal{P}}_s}$ increases with   $t_1$. When $t_1^+>t_1^*$, we have ${\cal{P}}_s^+>{\cal{P}}_s^*$, indicating that   $\left( {t_1^ * ,{t_2^ *},{\beta^* _{\rm{A}}} ,{\beta^* _{\rm{B}}},\lambda^ * } \right)$ is not the optimal solution.
\section{Proof of the Proposition 3}
We prove Proposition 3 from two steps as follows.  In {\emph{Step 1, }} we demonstrate that the maximum successful probability could be achieved when ${\max \left\{ {t_2^{\rm{A}},t_2^{\rm{B}},t_2^{{\rm{MEC}}}} \right\} = {t_2}}$ by contradiction. In \emph{Step 2,} we show that ${t_2^{\rm{A}}=t_2^{\rm{B}}=t_2^{{\rm{MEC}}}}$ holds for successful computation probability maximization.  For convenience, suppose that $\left( {t_1^ * ,{t_2^ *},{\beta^ *_{\rm{A}}},{\beta^ * _{\rm{B}}},\lambda^ * } \right)$ achieves the maximum successful computation probability ${\cal{P}}_s^*$ and that $\left( {t_1^ + ,{t_2^ +},{\beta^ + _{\rm{A}}},{\beta ^ +_{\rm{B}}},\lambda^ + } \right)$ is the other feasible solution for ${\bf{P}}_0$. The corresponding successful computation probability with the constructed solution is denoted as ${\cal{P}}_s^+$. We also assume that  $\left( {t_1^ * ,{t_2^ *},{\beta^ *_{\rm{A}}},{\beta^ * _{\rm{B}}},\lambda^ * } \right)  \ne \left( {t_1^ + ,{t_2^ +},{\beta^ + _{\rm{A}}},{\beta ^ +_{\rm{B}}},\lambda^ + } \right)$. Thus, we have ${{\cal{P}}^*_s}>{{\cal{P}}^+_s}$.

Proof of {\emph{Step 1.}} Suppose that $\max \left\{ {t_2^{\rm{A}},t_2^{\rm{B}},t_2^{{\rm{MEC}}}} \right\} <{t_2^ *}$ holds. It can be drawn from Proposition 2 that $t_1^*+t_2^*=T$. Also, it is assumed that $\lambda^ *=\lambda^ + $, $t_1^++t_2^+=T$, ${t_2^ +}=\max \left\{ {t_2^{\rm{A}},t_2^{\rm{B}},t_2^{{\rm{MEC}}}} \right\} $, $\beta^+_{\rm{A}}=\beta^*_{\rm{A}}$ and $\beta^+_{\rm{B}}=\beta^*_{\rm{B}}$ are satisfied.  Obviously, the constructed solution satisfies all the constraints of ${\bf{P}}_0$. Besides, since   $\beta^+_{\rm{A}}=\beta^*_{\rm{A}}$ and $\beta^+_{\rm{B}}=\beta^*_{\rm{B}}$ are satisfied, we have $t_2^+<t_2^*$ and $t_1^+>t_1^*$. As pointed out in the Proof of the Proposition 2, ${{\cal{P}}_s}$ increases with   $t_1$. Hence, we have ${{\cal{P}}^+_s}>{{\cal{P}}^*_s}$. This contradicts the original assumption that ${{\cal{P}}^*_s}>{{\cal{P}}^+_s}$. In summary, ${\max \left\{ {t_2^{\rm{A}},t_2^{\rm{B}},t_2^{{\rm{MEC}}}} \right\} = {t_2}}$ holds from successful computation probability maximization.

Proof of {\emph{Step 2.}}  We firstly introduce  Lemma 1 as follows.

\emph{Lemma 1.} For any given parameters satisfying all the constraints of $\bf{P}_0$, ${{\cal{P}}_s}$  decreases with the increase of $\beta_{\rm{A}}$ (or $\beta_{\rm{B}}$).

\emph{Proof.}  The first-order derivative of ${{\cal{P}}_s}$ with respect to ${\beta _{\rm{A}}}$ is written as
\begin{small}
\begin{align}
\frac{{\partial {{\cal{P}}_s}}}{{\partial {\beta _{\rm{A}}}}} = \frac{{\partial {{\cal{P}}_s}}}{{\partial {\gamma _1}}} \times \frac{{\partial {\gamma _1}_s}}{{\partial {\beta _{\rm{A}}}}} = \frac{{\partial {{\cal{P}}_s}}}{{\partial {\gamma _1}}} \times \frac{M}{{{t_1}{B_c}}}{2^{\frac{{{\beta _{\rm{A}}}M}}{{{t_1}{B_c}}}}}{\rm{ }}. \tag{C.1}
\end{align}
\end{small} Considering (B.2), it is easy to find that ${{\cal{P}}_s}$  decreases with the increase of $\beta_{\rm{A}}$. Similarly, we can also prove that ${{\cal{P}}_s}$  decreases with the increase of $\beta_{\rm{B}}$. \hfill {$\blacksquare $}

%It is quite straightforward to see that $\gamma_1$ (or $\gamma_2$) increases with  $\beta_{\rm{A}}$ (or $\beta_{\rm{A}}$). Thus, ${\frac{{{\gamma _1}\left( {1 + d_{{{\rm{A}}}}^\alpha } \right)\left( {{\rm{1 + }}{\gamma _2}} \right)}}{{\lambda \rho }}}$ and ${\frac{{1 + d_{{{\rm{B}}}}^\alpha {\gamma _2}}}{{\left( {1 - \lambda } \right)\rho }}}$  increase with $\beta_{\rm{A}}$ (or $\beta_{\rm{A}}$), while ${\frac{{\left( {1 + d_{{{\rm{B}}}}^\alpha } \right)\lambda \rho x - {\gamma _1}\left( {1 + d_{{{\rm{A}}}}^\alpha } \right)\left( {1 + d_{{{\rm{B}}}}^\alpha } \right)}}{{{\gamma _1}\left( {1 - \lambda } \right)\left( {1 + d_{{{\rm{A}}}}^\alpha } \right)\rho }}}$ decreases with the increase of  $\beta_{\rm{A}}$ (or $\beta_{\rm{A}}$). Together with (B.1), Lemma 1 can be proven. \hfill {$\blacksquare $}

Now we employ contradiction to verify that the maximum successful computation probability could be achieved when ${t_2^{\rm{A}}=t_2^{\rm{B}}=t_2^{{\rm{MEC}}}}=t_2$ holds. Assume that $t_2^{\rm{A}*}= t_2^{\rm{B}*} = t_2^{{\rm{MEC}*}}=t_2^*$ is not satisfied and  $t_2^{\rm{A}+}= t_2^{\rm{B}+} = t_2^{{\rm{MEC}+}}=t_2^+$ holds. We also assume that $\lambda^*=\lambda^+$ holds.
Obviously, the constructed solution satisfies all the constraints of ${\bf{P}}_0$. It can be derived from $t_2^{\rm{A}+}= t_2^{\rm{B}+} = t_2^{{\rm{MEC}+}}=t_2^+$ that ${t_2^ +  = \frac{{2MC}}{{{f_{{\rm{user}}}}\left( {N{\rm{ + 2}}} \right)}}}$ and $\beta _{\rm{A}}^ +  = \beta _{\rm{B}}^+ =1 - \frac{{{f_{{\rm{user}}}}t_2^ + }}{{MC}}$ are satisfied.
Based on the conclusion summarized in \emph{Step 1},  there are four
cases for the assumption that  $t_2^{\rm{A}*}= t_2^{\rm{B}*} = t_2^{{\rm{MEC}*}}=t_2^*$ is not satisfied as follows: Case 1 with $t_2^{\rm{A}*}=t_2^*$, $t_2^{\rm{B}*}\le t_2^*$ and $t_2^{{\rm{MEC}*}}<t_2$, Case 2 with $t_2^{\rm{A}*}=t_2^*$, $t_2^{\rm{B}*}<t_2$ and $t_2^{{\rm{MEC}*}}= t_2^*$, Case 3 with $t_2^{\rm{A}*}<t_2$, $t_2^{\rm{B}*}=t_2^*$  and $t_2^{{\rm{MEC}*}}\le t_2^*$ and Case 4 with $t_2^{\rm{A}*}<t_2$, $t_2^{\rm{B}*}<t_2^*$ and $t_2^{{\rm{MEC}*}}=t_2^*$ .

For Case 1, we have $\beta _{\rm{A}}^* = 1 - \frac{{{f_{{\rm{user}}}}t_2^*}}{{MC}}$, $\beta _{\rm{B}} \ge 1 - \frac{{{f_{{\rm{user}}}}t_2^*}}{{MC}}$ and $\beta _{\rm{A}}^* + \beta _{\rm{B}} < \frac{{2N{f_{{\rm{user}}}}t_2^*}}{{MC}}$. Thus, $2 - \frac{{2{f_{{\rm{user}}}}t_2^*}}{{MC}} \le \beta _{\rm{A}}^* + \beta _{\rm{B}}^* < \frac{{2N{f_{{\rm{user}}}}t_2^*}}{{MC}}$ and $1 - \frac{{{f_{{\rm{user}}}}t_2^*}}{{MC}} < \frac{{N{f_{{\rm{user}}}}t_2^*}}{{MC}}$ are satisfied. Combing Lemma 1, $\beta _{\rm{A}}^*$ and $\beta _{\rm{B}}^*$ are determined, i.e., $\beta _{\rm{A}}^* = \beta _{\rm{B}}^* = 1 - \frac{{{f_{{\rm{user}}}}t_2^*}}{{MC}}$. Besides, it can be inferred from $1 - \frac{{{f_{{\rm{user}}}}t_2^*}}{{MC}} < \frac{{N{f_{{\rm{user}}}}t_2^*}}{{MC}}$ that $t_2^* > \frac{{MC}}{{{f_{{\rm{user}}}}\left( {N + 1} \right)}}$ holds. Since $t_2^*>t_2^+$, we have $\beta _{\rm{A}}^*<\beta _{\rm{A}}^+$ and $\beta _{\rm{B}}^*<\beta _{\rm{B}}^+$.   Based on Proposition 2 and Lemma 1, as well as the fact that ${{\cal{P}}_s}$ increases with   $t_1$, the following inequalities in (C.2) are satisfied and thus Case 1 is  not optimal.
\begin{small}
\begin{align}\notag
{\cal{P}}_s^ + &={\cal{P}}_s^{}\left( {t_1^ + ,T - t_1^ + ,\beta _{\rm{A}}^ + ,\beta _{\rm{B}}^ + ,{\lambda ^*}} \right) > {\cal{P}}_s^{}\left( {t_1^*,T - t_1^*,\beta _{\rm{A}}^ + ,\beta _{\rm{B}}^ + ,{\lambda ^*}} \right)\\ \notag
&> {\cal{P}}_s^{}\left( {t_1^*,T - t_1^*,\beta _{\rm{A}}^ + ,\beta _{\rm{B}}^ + ,{\lambda ^*}} \right) > {\cal{P}}_s^{}\left( {t_1^*,T - t_1^*,\beta _{\rm{A}}^*,\beta _{\rm{B}}^ + ,{\lambda ^*}} \right) \\ \notag
&> {\cal{P}}_s^{}\left( {t_1^*,T - t_1^*,\beta _{\rm{A}}^*,\beta _{\rm{B}}^*,{\lambda ^*}} \right) = {\cal{P}}_s^{}\left( {t_1^*,t_2^*,\beta _{\rm{A}}^*,\beta _{\rm{B}}^*,{\lambda ^*}} \right) = {\cal{P}}_s^*.\tag{C.2}
\end{align}
\end{small}By the same argument, the fact that Cases 2-4 are also not satisfied can be proven readily and the proof is omitted due to the space limitation.
\section{Proof of the Theorem 2}
In order to obtain the optimal power allocation coefficient $\lambda^* $, we firstly study the first-order derivative of ${{\cal{P}}_s}$ with respect to $\lambda$, given by {\small{$\frac{{\partial {{\cal{P}}_s}}}{{\partial \lambda }} = \frac{{\exp \left( { - \frac{{{a_2}}}{{1 - \lambda }} - \frac{{{a_3}}}{\lambda }} \right)}}{{\lambda  + {a_1}\left( {1 - \lambda } \right)}}\left( {\frac{{{a_1}}}{{\left( {{\rm{1}} - {a_1}} \right)\lambda  + {a_1}}} + \frac{{{a_3}}}{\lambda } - \frac{{{a_2}\lambda }}{{{{\left( {1 - \lambda } \right)}^2}}}} \right)$}}.
Obviously, the sign of $\frac{{\partial {{\cal{P}}_s}}}{{\partial \lambda }}$ corresponds with $\Xi={\left( {\frac{{{a_1}}}{{\left( {{\rm{1}} - {a_1}} \right)\lambda  + {a_1}}} + \frac{{{a_3}}}{\lambda } - \frac{{{a_2}\lambda }}{{{{\left( {1 - \lambda } \right)}^2}}}} \right)} $, and thereby we can obtain  $\lambda^* $ by solving $\Xi =0$, which can be written as
\begin{small}
\begin{align}
 {\frac{{{a_1}}}{{\left( {{\rm{1}} - {a_1}} \right)\lambda  + {a_1}}}} =  {\frac{{{a_2}\lambda }}{{{{\left( {1 - \lambda } \right)}^2}}} - \frac{{{a_3}}}{\lambda }}.\tag{D.1}
\end{align}
\end{small}Constrained as $\lambda  \in \left( {0,1} \right)$,  ${g\left( \lambda  \right)}=\frac{{{a_2}\lambda }}{{{{\left( {1 - \lambda } \right)}^2}}} - \frac{{{a_3}}}{\lambda }$ increases with $\lambda$  and the range of ${g\left( \lambda  \right)}$ is $\left( { - \infty , + \infty } \right)$; ${f\left( \lambda  \right)}={\frac{{{a_1}}}{{\left( {{\rm{1}} - {a_1}} \right)\lambda  + {a_1}}}}$ with $a_1>1$ increases (or with $a_1<1$ decreases) as we increase $\lambda$, while ${f\left( \lambda  \right)}=1$ when $a_1=1$.  In other words, within the feasible region, ${f\left( \lambda  \right)}$ with $a_1>1$ (or $a_1<1$) monotonically increases (or decreases) from 1 to  $a_1$, and ${g\left( \lambda  \right)}$ monotonically decreases from positive infinity to negative infinity. Thus, there exists a unique ${\lambda ^*} \in \left( {0,1} \right)$ with satisfying (D.1), and the ${\lambda ^*}$ can be obtained by using simple iterative algorithms. Nevertheless, we still derive the closed-form expression for ${\lambda ^*}$ as follows. Through mathematical calculations we rewrite (D.1) as
\begin{align}
{m_1}{\lambda ^3} + {m_2}{\lambda ^2} + {m_3}\lambda  + {a_1}{a_3} = 0.\tag{D.2}
\end{align}
According to \cite{nickalls1993new}, there are three cases for the solution of (D.2) based on the value of $y_N$, i.e.,
{Case i.} If ${y^2_N} - {g^2} > 0$, there exists a real root and we have
\begin{small}
\begin{align}
\lambda^* \! =\! {x_N} \!+\! \sqrt[3]{{\frac{{ - {y_N} + \sqrt {{y_N^2} - {g^2}} }}{{2{m_1}}}}} \!+\! \sqrt[3]{{\frac{{ - {y_N} \!-\! \sqrt {{y_N^2} - {g^2}} }}{{2{m_1}}}}}.\tag{D.3}
\end{align}
\end{small}{Case ii.} If ${y^2_N} - {g^2} = 0$, there exist two real roots,
\begin{small}
\begin{align}
{\lambda ^*} = \left\{ {0 < {\lambda _i} < 1\left| {{\lambda _i} = {x_N} + \delta ,\;{\lambda _i} = {x_N} - 2\delta } \right.} \right\}.\tag{D.4}
\end{align}
\end{small}{Case iii.} If ${y^2_N} - {g^2} < 0$, there exist  three roots,
\begin{small}
\begin{align}\notag
\!\!\!\!{\lambda ^*} \!=\! \left\{ \!{0\! <\! {\lambda _i}\! <\! 1\left| {{\lambda _i}\! =\! {x_N} \!+\! 2\delta \cos \left( {\phi\!  -\! \frac{{2\pi i}}{3}}\! \right),i\! =\! 0,1,2} \right.} \right\}.\tag{D.5}
\end{align}
\end{small}%Combing Cases i$-$iii, Theorem 2 is proven.
 \end{appendices}

\vspace{-20pt}
\ifCLASSOPTIONcaptionsoff
  \newpage
\fi
\bibliographystyle{IEEEtran}
\bibliography{ref}
\end{document}